\documentclass[twocolumn,showpacs,preprintnumbers,amsmath,amssymb]{revtex4}

\usepackage{amsmath}
\usepackage{graphicx}
\usepackage{color}
\usepackage{dcolumn}
\usepackage{bm}

\def\slap{p \hspace{-1.7mm} \slash}

\begin{document}

\preprint{DCP-08-01}

\title{Discrete Higgs and the Cosmological Constant}

\author{
Paolo Amore,$^{1}$\footnote{Electronic address:paolo.amore@gmail.com}
Alfredo Aranda,$^{1,2}$\footnote{Electronic address:fefo@ucol.mx} 
and J. L. Diaz-Cruz$^{2,3}$\footnote{Electronic address:ldiaz@sirio.ifuap.buap.mx}} 

\affiliation{$^1$Facultad de Ciencias,CUICBAS,  Universidad de Colima,\\
  Bernal D\'{i}az del Castillo 340, Colima, Colima, M\'exico \\
  $^2$Dual C-P Institute of High Energy Physics \\
  $^3$C.A. de Part\'iculas, Campos y Relatividad \\
  FCFM-BUAP, Puebla, Pue. M\'exico}

\date{\today}

\begin{abstract}
  It is proposed that the Higgs vacuum possesses a small-scale structure that can explain
the large discrepancy between the predicted electroweak vacuum energy density and
the observed cosmological constant. An effective Lagrangian description is employed to
obtain modifications to the Standard Model predictions that can be tested at collider 
experiments.
\end{abstract}

\pacs{}

\maketitle

It is expected that the start up of the Large Hadron Collider (LHC) will open up the window
for the detailed study of electroweak scale physics, leading to a better understanding of 
the actual process of electroweak symmetry breaking (EWSB), and the generation of mass for the quarks, 
leptons and gauge bosons of the Standard Model (SM)~\cite{lhc1,lhc2}. Understanding the physics associated to
this problem is paramount for the development of particle physics, for it represents the seed 
in all of the so-called physics beyond the SM.

A well known problem associated with EWSB pertains to the contribution that this breaking gives 
to the vacuum energy~\cite{Weinberg:1988cp}. In its simplest version, that of the SM, 
EWSB occurs through the spontaneous
breaking of the symmetry via the presence of a scalar field, the Higgs, whose vacuum expectation value 
(vev) breaks SU(2)$_W \times$U(1)$_Y$ down to U(1)$_{QED}$ rendering the SM gauge bosons $Z$ and $W^{\pm}$
massive. The experimental determination for the $W$ mass gives the value for the Higgs vev $v=246$~GeV.
The Higgs potential in the SM is given by
\begin{eqnarray}
  \label{potentialSM}
  V(\Phi^{\dagger}\Phi) = -\mu^2\Phi^{\dagger}\Phi + \lambda (\Phi^{\dagger}\Phi)^2 \ ,
\end{eqnarray}
where $\Phi$ is an SU(2)$_W$ doublet. When the Higgs acquires its vev and spontaneously breaks electroweak
symmetry, it also contributes to the vacuum energy by a factor of order 
$\Lambda_{EW} \sim \lambda (\langle \Phi^{\dagger}\Phi \rangle)^2 \sim v^4 \sim 10^9$~GeV$^4$. 
This is to be contrasted with the current observed value for the cosmological 
constant $\sim 10^{-47}$~GeV$^4$~\cite{Knop:2003iy}. Other contributions to the cosmological constant 
include the vacuum condensate associated with chiral symmetry breaking in QCD, as well as
the vacuum fluctuations associated with the zero-point energies of quantum
fields, both of which appear to be too large by many orders of magnitude.
This problem is present in all of the extensions 
beyond the SM which have
quantum field theory as their underlying structure. Solution to this conundrum has been at best postponed 
hoping that perhaps a better understanding of gravity at the quantum level might explain it. For a
complete review of this situation see~\cite{Bousso:2007gp}.

This letter describes an idea that can be of relevance for both the EWSB sector and its possible contribution
to the vacuum energy. It is based on the hypothesis that the Higgs vacuum is not uniform in space but rather 
has an inhomogeneous structure. As a working example the vacuum is considered to contain vev-filled spherical 
droplets (of radius $r_d$) distributed in a regular array with inter-droplet separation $l_d$. 
Denoting by $\tau_d$ the droplet volume, the contribution to the vacuum energy  is estimated to be 
$\Lambda_{d}\simeq \rho_d \Lambda_{EW}\tau_d$, where $\rho_d$ is the droplet density. Saturating the observed
value yields $\rho_d\tau_d \sim 10^{-56}$. 

Since this structure has not been observed, $l_d$ (and hence $r_d$)
must be smaller than the current explored distance: $l_d \leq 10^{-15}$~cm. Again saturating this
constraint (assuming $l_d \sim 10^{-15}$~cm) results in the following estimate:
\begin{eqnarray}
  \label{dropletsize} \nonumber
  \rho_d\tau_d & \sim & \frac{r_d^3}{l_d^3} \sim 10^{-56} \\
  & \rightarrow & r_d \sim 10^{-33} \ \rm{cm} \ .
\end{eqnarray}

It is indicative that in this simple scenario, assuming only that the characteristic inter-droplet
distance is O($10^{-15}$~cm), the droplet size turns out to be of order $l_{\rm{Planck}}=10^{-33}$~cm. 

\begin{figure}[ht]
  \begin{center}
    \includegraphics[width=5cm]{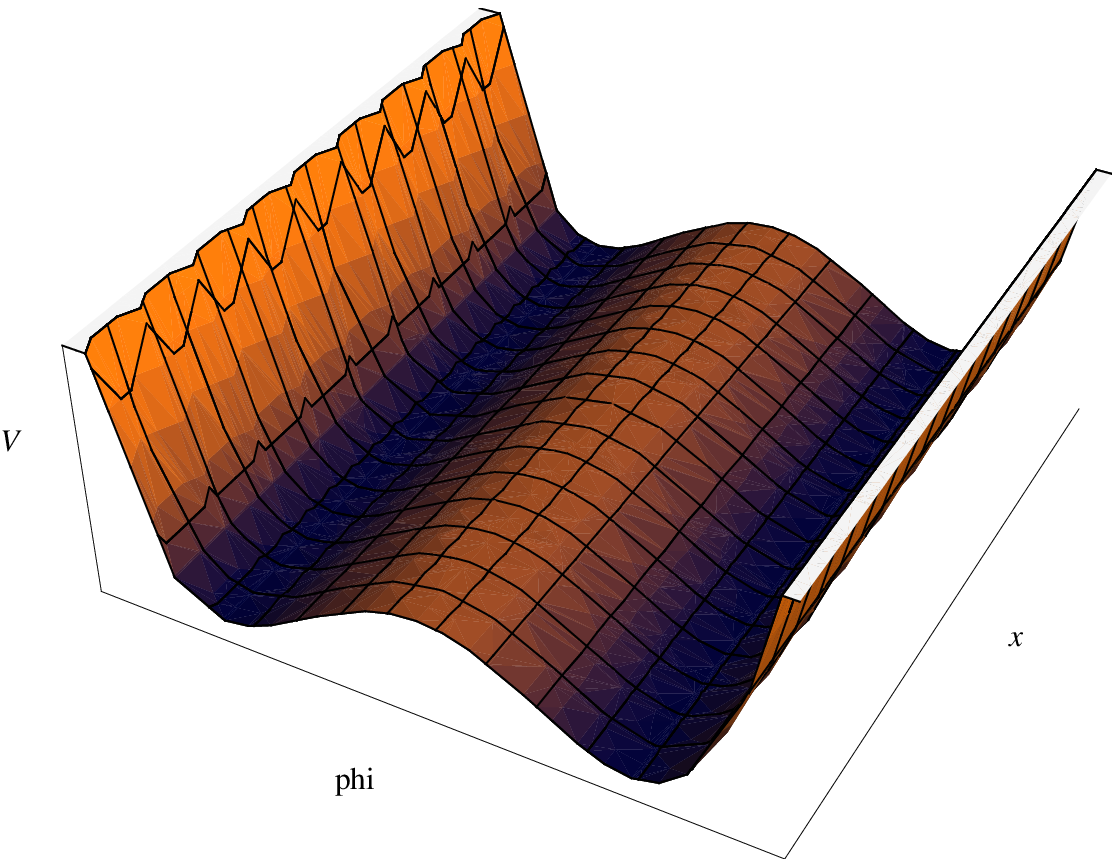}
    \includegraphics[width=5cm]{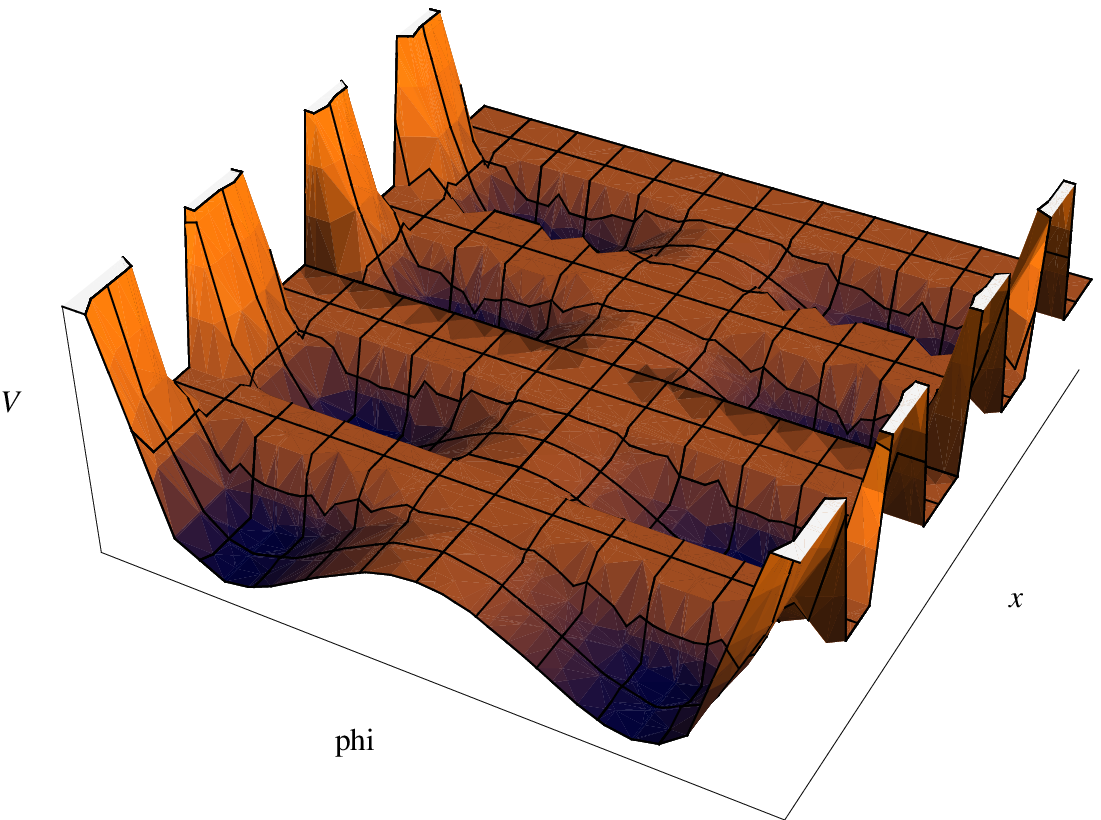}
    \caption{Schematic view of the Higgs potential in the $\phi - x$ plane for the case of the Standard Model
      (top) and for uniform spherical droplets (bottom).}
    \label{fig:potential}
  \end{center}
\end{figure}

In terms of the Higgs potential this scenario represents a case where the vev is not uniform over spacetime. 
The simple model above is represented in Figure~\ref{fig:potential} (bottom) where the potential is shown in the
$\phi - x$ plane. Also shown is a case that corresponds to the SM.

In order to explore the possibility of seeing an experimental effect at colliders due to this small
scale structure of 
the vacuum, under this example's assumptions, the following analysis is presented:

Consider a probe of wavelength $\lambda_p$. Then there are three different relevant scales: i) the scale where
$\lambda_p \gg l_d$ with quantum field theory and massive particles. In this region the probe feels 
a broken SU(2)$_W \times$ U(1)$_Y \rightarrow$ U(1)$_{QED}$ with vev $v=246$~GeV (interpreted as the
average of the droplet distribution), ii) the region where $\lambda_p \sim l_d$ and the effects
of the small scale structure will start showing up. This region is described within the framework of
quantum field theory by an effective Lagrangian ${\cal{L}}_{eff}$, and iii) the region where 
$\lambda_p \sim r_d$ where a new quantum vacuum dynamics takes place and ultimately determines the shape, 
distribution and physical description of the droplets (particles are massless in this region). 

In summary

\begin{itemize}
\item $\lambda_p \gg l_d \rightarrow$ QFT and massive particles
\item $\lambda_p \sim l_d \gg r_d \rightarrow$ ${\cal{L}}_{eff}$
\item $\lambda_p \sim r_d \rightarrow$ New quantum vacuum dynamics
\end{itemize}

The first two regions above can be described using the language of QFT through an effective Lagrangian
that reduces to the SM Lagrangian when $\lambda_p \gg l_d$. One way to accomplish this is to
parametrize the new unknown effects due to the vacuum's small scale structure, i.e. 
the ignorance of the quantum vacuum dynamics, into the metric in the following way:
\begin{eqnarray} 
  \label{metric}
  G_{\mu \nu} = a(E) g_{\mu \nu} = (1+\eta(E)) g_{\mu \nu} \ ,
\end{eqnarray}
where $a(E) = 1+\eta(E)$ with $\eta(E) \ll 1$ for $E<1/l_d$. This is motivated by the idea that
once the vacuum structure is perceived by the particles, their masses and dispersion relations
will be affected by it and thus, Lorentz invariance will be lost. In this simple example it is assumed that
all entries in the metric are modified by the same factor $a(E)$. This is certainly an oversimplification 
and more complex scenarios will be investigated. However, even this simple
setting leads to possible physical effects and it is presented to exemplify the general idea.
Using this modification then leads to the general product 
$\hat{A}\hat{B} \equiv \hat{A}_{\mu}\hat{B}^{\mu}=a(E)A_{\mu}B^{\mu}$ and the 
factor $a(E)$ then feeds into the propagators and Feynman rules of the SM. 

The expressions for the propagators are:
\begin{widetext}
  \begin{eqnarray}
    \label{propagators} \nonumber
    \frac{i}{\hat{p}^2-m^2}  =  \frac{i}{a(E)p^2-m^2}  & ,& \ \ \
    \frac{i(\hat{\slap}-m)}{\hat{\slap}\hat{\slap}-m^2}   =   
    \frac{i\left(a(E)\slap-m\right)}{a(E)^2 p^2-m^2} \ , \\ 
    \frac{-i}{\hat{q}^2-\hat{m}_V^2} \times  \left(G^{\mu \nu} -
    \frac{\hat{q}^{\mu}\hat{q}^{\nu}}{\hat{m}_V^2}\right)  & = & 
    \frac{-i}{q^2-m_V^2}  \times  \left(g^{\mu \nu} - \frac{q^{\mu}q^{\nu}}{a(E)m_V^2}\right) \ .
  \end{eqnarray}
\end{widetext}

Note that the mass terms in the massive vector boson propagator contain a factor of $a(E)$. This is due
to the fact that the mass terms come from the $\hat{A}^{\mu}\hat{A}_{\mu}$ in the Lagrangian. Note also that
the only propagator that does not receive a modification in this case is that of a massless vector boson,
an expected result due to the metric Eq.(\ref{metric}).

The physically observable implications of this kind of scenarios at colliders are then obtained by finding
the specific deviation from the SM predictions. Consider for example the $Z$ width, which in the example
of this letter turns out to be $\Gamma_Z=a(e)^3 \Gamma_{Z}^0 \approx \Gamma_{Z}^0 (1+3\eta(E))$, where
$\Gamma_Z^0$ denotes the SM expression. The previous result was obtained using the fact that the usual term
in the SM Lagrangian involving the $Z$ boson was modified as $\hat{Z}^{\mu}\hat{{\cal{J}}}_{\mu}
\rightarrow a(E)Z^{\mu}{\cal{J}}_{\mu}$ and the vector polarization sum for the external $Z$ was taken to be
$\sum_i \hat{\epsilon}(q)_{\mu}^{(i)}\hat{\epsilon}(q)_{\nu}^{(i)*}=-a(E)g_{\mu \nu}+q_{\mu}q_{\nu}/m_Z^2$.

Using the previous result leads to the following expression for the process $e^+ \ e^- \to f \ \bar{f}$
at $\sqrt{s}=m_Z$:
\begin{eqnarray}
  \label{drellyan}
  \sigma(e^+e^-\to f\bar{f})_{\rm{peak}} \approx \left(1 - 2\eta(E)\right)\sigma(e^+e^-\to f\bar{f})_{\rm{peak}}^0  \ ,
\end{eqnarray}
where again $\sigma(e^+e^-\to f\bar{f})_{\rm{peak}}^0$ stands for the SM expression~\cite{Ellis:1991qj}. Given that
LEP-I reached a precision of O($0.1\%$) in the Z-width determination~\cite{PDG}, this can be translated into the 
constraint $\eta \sim 5 \times 10^{-4}$. 

A full analysis involving the scalar sector of the SM and precision tests is necessary in order to confront
this type of scenario with experiments. In the simple case above the scalar-vector interactions are given by
\begin{eqnarray}
  \label{lhvv}
  {\cal{L}}_{hVV} = g_{hV}m_W h \hat{V}_{\mu}\hat{V}^{\mu} = a(E) g_{hV} m_W  h V_{\mu}V^{\mu} \ ,
\end{eqnarray}
where $g_{hW}=g$, and $g_{hZ}=g/\cos^2(\theta_W)$. This leads to the following expression for the Higgs decays
to ZZ and WW (to leading order in $\eta$):
\begin{eqnarray}
  \label{hwidth}
  \Gamma(h \to WW) & = & \Gamma^{hWW}_{SM} + \eta \frac{3 g^2 m_h x_W}{64 \pi c_w^4} \sqrt{x_W-1}  \ , \\
  \Gamma(h \to ZZ) & = & \Gamma^{hZZ}_{SM} + \eta \frac{3 g^2 m_h x_Z}{128 \pi c_w^4} \sqrt{x_Z-1}  \ ,
\end{eqnarray}
where $\Gamma^{hVV}_{SM}$ stands for the SM expressions and $x_V \equiv 4 m_V^2/m_h^2$. Taking into account
the fact that at the LHC these widths could be determined to the $10 - 20\%$ level~\cite{Assamagan:2004mu}
imposes the constraint $\eta \le$ O($10^{-1}$), which is weaker than the constraint above.

Parametrizing the unknown quantum vacuum dynamics that characterizes
this setup in full generality, i.e. using $G_{\mu \nu} = g_{\mu \nu} + \Delta_{\mu \nu}$, will certainly
lead to interesting effects not present in the simple example explored in this letter. That analysis is currently
being pursued. 

The purpose of this letter is to show that by considering a small scale structure of the vacuum, that is,
taking the Higgs vev to be spacetime dependent at some high energy scale, it is possible to propose a solution to the 
vacuum energy contribution due to EWSB, and at the same time render observable effects at collider
experiments. It is remarkable that in the simple model where the vacuum is characterized by a uniform
distribution of vev-filled spherical droplets, and imposing the condition that the inter-droplet separation is of 
the order of the smallest explored distance, leads to a droplet size of Planckian length automatically. 

A recent discussion by Brodsky and Shrock~\cite{Brodsky:2008xu} proposes a solution to the QCD contribution to 
$\Lambda$ similar in spirit to the considerations presented in this letter, and a link between the Higgs and
dark matter is discussed in~\cite{DiazCruz:2007be}

\begin{acknowledgments}
The authors acknowledge support from CONACyT and SNI. AA thanks the Facultad de Ciencias F\'isico-Matem\'aticas - BUAP
for their hospitality while part of this work was being done. 
\end{acknowledgments}

\end{document}